\documentclass[a4paper,showpacs,amsmath,amssymb,12pt]{article}

\usepackage{mathrsfs}
\usepackage{amsfonts,color,slashbox}
\usepackage{indentfirst}
\usepackage{amssymb}
\usepackage{amsmath}
\usepackage[dvips]{graphicx}

\numberwithin{equation}{section}
\title{Tiny Graviton Matrix Theory On Time-Dependent Background}
\author{Bin Chen and Xiao Liu\footnote{Email:bchen01,liuxiaoerty@pku.edu.cn}\\
{\small Department of Physics} \\
{\small and State Key Laboratory of Nuclear Physics and Technology,}\\
{\small Peking University, Beijing 100871, P.R.China}}
\date{}

\begin{document}

\maketitle

\begin{abstract}

In this article we construct a tiny graviton matrix model for type
IIB string theory on a plane-wave background with null dilaton. For
the linear null dilaton case, we analyze its vacuum and the
excitation spectrum around the vacuum, and discuss the
time-dependent fuzzy three-sphere solutions and their evolution. It
turns out that at very late time the non-Abelian fuzzy degrees of
freedom disappear, which indicates the appearance of perturbative
strings.

\end{abstract}

\newpage

\section{Introduction}

One of central problems in the study of string theory is to look for
its non-perturbative definition. One decade ago, Banks, Fischler,
Shenker and Susskind (BFSS) suggested that a matrix model could be a
candidate  for non-perturbative definition of string theory (M
theory)\cite{bfss}. They constructed so-called BFSS matrix model on
the flat background, and conjectured that large $N$ limit of their
model is equivalent to uncompactified
%discrete
%light-cone quantization (DLCQ) of
M theory in the infinite momentum frame. In \cite{sei,sen}, the
authors showed how to obtain BFSS model from discrete light-cone
quantization (DLCQ) of
M theory. %they argued that when $N$
%is finite, BFSS matrix model describe DLCQ of M theory correctly.
%via DLCQ method.
The BFSS model is a $(0+1)$-dimensional super-Yang-Mills gauge
theory, or a matrix model, and its Hamiltonian can be obtained
from the nonrelativistic Hamiltonian of D-particles. Remarkably,
there is another way to obtain this model by doing matrix
regularization of super-membrane in flat space-time.

Since the construction of BFSS
model, %After BFSS's conjecture, lots of work about matrix model
%was done.
people has tried hard to construct the matrix models on general
backgrounds. For the weakly curved spacetime,  the authors in
\cite{tay1} studied the non-Abelian D-particle action, which could
be
essential to get the matrix model action. %is an example. Those
%matrix models are all background-dependent.
Even for this case, the complete D-particle action is still not
clear partially due to notorious difficulty in constructing
non-Abelian
D-brane action. % It is still a non-completed work to construct
%matrix model on all backgrounds.
However, for one class of curved spacetime, so-called plane-wave
background, the corresponding BMN matrix model was well-established
and much better
understood\cite{bmn}. %An important example of those matrix models
%is the so-called BMN matrix model associated to the pp-wave
%background\cite{bmn}.
%, pointed out by D.Berenstein, J.Maldcena and
%H.Nastase (BMN) .
Similar to the matrix model in flat spacetime, the BMN model could
be obtained from the matrix regularization of supermembrane on the
background\cite{DSR2002}. This suggest that the matrix
regularization of super-membrane could be an important way to
construct the matrix model in curved spacetime. For the general
curved spacetime, the action of the supermembrane could not be
determined completely\cite{Peeters1998}. However, it is
remarkable, for a general class of plane-wave-like background, the
matrix model could be constructed to all order\cite{chen2}.
%For time-dependent
%background, there are several matrix models about them
%too\cite{chen2}.

%Matrix model usually associated to M theory.
There are also other nonperturbative definition of various kinds of
string theory. For IIA theory, one non-perturbative definition of
string theory is matrix string theory\cite{DVV97}. %some matrix models also
%were constructed.
%One class of these models is
%matrix membrane model. It is a $0+2$-dimensional matrix model.
For IIB theory, one interesting construction is the IKKT matrix
model \cite{ikkt}, which is a $(0+0)$-dimensional theory. Quite
recently, a $(0+1)$-dimensional matrix model associated to IIB
theory on plane-wave background was proposed by
Sheikh-Jabbari\cite{jabatiny}\footnote{For an alternative proposal
to the Tiny graviton matrix theory, see \cite{Lozano}.}. His
observation is that the matrix models, including BFSS and BMN, can
be understood as the theory of gravitons. In BFSS case, the
D-particles are gravity waves in 11-dimensional flat spacetime,
while in BMN case, the fundamental degrees of freedom are tiny
graviton, which are membrane gravitons with very small size. In
other words, the DLCQ of M-theory in 11-dimensional background could
be described by a theory of $N$ membrane graviton. This led
Sheikh-Jabbari to conjecture that
%Jabbari analyze tiny graviton
%\cite{jabatiny} on plane-wave background, found that $R_{tiny}\ll
%l_{pl}$.  More than this, tiny graviton is a half BPS state. And
%when J tiny gravitons are coincident, $U(J)$ gauge symmetry will
%emerge.  These make tiny graviton looks like D$0$-brane.
%Moreover, one tiny graviton is associated with one unit of compact
%momentum. Comparing with BFSS and BMN matrix model, Jabbari
DLCQ of IIB superstring theory on plane-wave background is a
$0+1$-dimensional matrix model, whose Hamiltonian can be written
down through the action of tiny gravitons. In this case the tiny
gravitons are very small spherical D3-branes, so  the Hamiltonian
can be obtained from light-cone matrix regularization and
quantization of D$3$-brane Hamiltonian on the background. More study
about tiny graviton matrix model can be found in
\cite{jabatiny2,jabatiny3,jabatiny4}.

The search for a non-perturbative definition of string theory is not
only just to complete the formulation of the theory, but may also
has profound physical applications. In particular, at the very early
universe,  the appropriate description of the physics there could be
in terms of nonperturbative string degrees of freedom since the
string coupling could be strong. One concrete realization of this
idea is so called matrix big bang model and its cousins. The
original matrix big bang is starting from the IIA string theory in
null dilaton background\cite{verlinde}. Formally, this IIA string
theory is exactly solvable perturbatively but since the string
coupling becomes strong at null infinity, the perturbative string
theory breaks down. It turned out that the matrix degrees of
freedom, rather than the point particle or the perturbative string,
could describe the physics near the null singularity. Following the
prescription in \cite{sei,sen}, a dual matrix string theory was
proposed. The study of the plane-wave matrix big bang model via DLCQ
was carried out very recently\cite{Blau1}. For a nice review on this
model, its generalization and relevant references, please see
\cite{Craps}.

%Time-dependent background was studied in many ways. Among these,
%some background can make string theory $\sigma$-model exact
%solvable.
In \cite{chen}, one class of solvable IIB string theory in linear
null dilaton background has been investigated. The background is %they found
%such a background for type IIB superstring theory .
\begin{eqnarray}\label{background}
&&ds^{2}=-2dx^{
+}dx^{-}-\lambda(x^{+})x^{2}_{I}dx^{+}dx^{+}+dx^{I}dx^{I},
\nonumber\\
&&\phi=\phi(x^{+}),
\nonumber\\
&&(F_{5})_{+1234}=(F_{5})_{+5678}=4f(x^{+}),
\end{eqnarray}
in which one has to require
\begin{eqnarray}\label{lamdef}
\lambda(x^{+})=-\frac{1}{4}\phi''(x^{+})+f^{2}(x^{+})e^{2\phi(x^{+})},
\end{eqnarray}
to satisfy the equations of motion.  This background keeps sixteen
non-linearly realized supersymmetries. The background has continuous
symmetry algebra $so(4)\oplus so(4)\oplus_s h(8)$, in which two
$so(4)$ are the residual rotational symmetry algebra and $h(8)$ is
the Heisenberg-type algebra.

Generically the background (\ref{background}) has curvature
singularity. To show this clear, it is better to work in the
Einstein frame. For the  linear null dilaton case we are interested
in this paper,
\begin{equation}
\phi =-c x^+, \hspace{5ex}\mbox{with $c>0$}
\end{equation}
the issue has been discussed carefully in section two of
\cite{chen}. It has been shown that no matter what the background
flux is the background (\ref{background}) is geodesically incomplete
and $x^+ \to -\infty$ is a curvature singularity where an inertial
observer will feel a divergent tidal force. The existence of
singularity and  the fact that the string coupling is also divergent
at $x^+ \to -\infty$ both suggest that the perturbative string
degrees of freedom to which the graviton belong can not give correct
description of the physics there, and instead nonperturbative string
will take over and play a fundamental role. In other words, to
describe the physics near $x^+ \to -\infty$ calls for a
nonperturbative definition of string theory, which might be a matrix
model.

 Similar to the usual plane-wave background, one can do
light-cone quantization of perturbative string in the background
(\ref{background}). However, due to the absence of translational
Killing vector along $x^{+}$, the lightcone Hamiltonian is not
conserved. Actually both the Hamiltonian and perturbative string
states are time-dependent. Nevertheless, because this background is
invariant under translation in the $x^{-}$ direction, the
compactification of $x^{-}$ on a circle gives a conserved lightcone
momentum. This fact also suggests that DLCQ
 of string and D-brane in this background is feasible. In this paper,
we will follow the proposal in \cite{jabatiny} to construct a
time-dependent matrix model on the background (\ref{background}).

% Then  Their early
%time and late time behaviors are our main concerns.

This article is organized as follows. In section 2, after
constructing the matrix model, we  analyze the spectrum around the
vacuum, and compare its spectrum with the spectrum of perturbative
string on this background. In section 3, we  look for and discuss
the fuzzy three sphere solutions. In our cases, the fuzzy
three-sphere solutions are time-dependent, with their radius
varying with time.  At last, we end with conclusions and some
discussions. In the Appendix, we discuss the giant graviton in the
background and show that its dynamics is the same as that of fuzzy
three sphere.

\section{The matrix model}

To construct the matrix model, we follow the construction proposed
in \cite{jabatiny}. Let us start from the light-cone Hamiltonian of
D$3$-brane. The Born-Infeld and Chern-Simons action for D$3$-brane
is\footnote{We will always set $l_{s}=1$ in this paper.}
\begin{equation}
  S=\int d\tau d^{3}\sigma
  e^{-\phi}\sqrt{-\textrm{det}(G_{\mu\nu}\partial_{\mu'}X^{\mu}\partial_{\nu'}X^{\nu})}+\int
  C_{4}.
\end{equation}
Here $G_{\mu\nu}$, $C_{4}$ are the spacetime metric and
Remond-Remond four-form potential of the background
(\ref{background}).  The $X^{\mu}$'s are the embedding coordinates.
The indices $\mu',\nu'=0,1,2,3$ label the D-brane's world-volume
coordinates,  and $i',j',\ldots$  denote the spatial ones.

In the light-cone gauge,
\begin{equation}
 X^{+}=\tau,\    G_{\mu\nu}\partial_{0}X^{\mu}\partial_{i'}X^{\nu}=0.
 \end{equation}
We use $P_{-}$ to denote the conjugate momentum of $X^{-}$,
$P^{+}=-P_{-}$, and $P_{I}$ to denote the conjugate momenta of
$X^{I}$, $I=1,2,\ldots,8$. Using the light-cone Hamiltonian
density $\mathcal
{H}=P_{-}\partial_{\tau}X^{-}+P_{I}\partial_{\tau}X^{I}-\mathcal
{L}$, we have the bosonic part of light-cone Hamiltonian:
\begin{eqnarray}
H_{bos}&=&\int
d^{3}\sigma[\frac{1}{2P^{+}}(P^{2}_{i}+P^{2}_{a})+\frac{\lambda(\tau)
P^{+}}{2}(X^{2}_{i}+X^{2}_{a})
\nonumber\\
&&\
+\frac{e^{-2\phi(\tau)}}{2P^{+}}(\frac{1}{3!}\{X^{i},X^{j},X^{k}\}\{X^{i},X^{j},X^{k}\}
\nonumber\\
&&\ +\frac{1}{3!}\{X^{a},X^{b},X^{c}\}\{X^{a},X^{b},X^{c}\}
\nonumber\\
&&\ +\frac{1}{2!}\{X^{i},X^{j},X^{a}\}\{X^{i},X^{j},X^{a}\}
+\frac{1}{2!}\{X^{i},X^{a},X^{b}\}\{X^{i},X^{a},X^{b}\})
\nonumber\\
&&\
-\frac{f(\tau)}{6}(\epsilon^{ijkl}X^{i}\{X^{j},X^{k},X^{l}\}+\epsilon^{abcd}X^{a}\{X^{b},X^{c},X^{d}\})],
\end{eqnarray}
where $i,j,\cdots=1,2,3,4$, $a,b,\cdots=5,6,7,8$. In the above
relation the Nambu three brackets are used,
\begin{eqnarray}\{X^{I},X^{J},X^{K}\}\equiv\epsilon^{i'j'k'}\partial_{i'}X^{I}\partial_{j'}X^{J}\partial_{k'}X^{K}.
\end{eqnarray}

The fermionic part of the Hamiltonian can be obtained through
superspace techniques\cite {d3brane,superspace,jabahami}. On the
background (\ref{background}), it is given by
\begin{eqnarray}
H_{fer}&=&\int
d^{3}\sigma[fe^{\phi}\psi^{\dagger\alpha\beta}\psi_{\alpha\beta}
\nonumber\\
&&\ +\frac{2e^{-\phi}}{P^{+}}
(\psi^{\dagger\alpha\beta}(\sigma^{ij})^{\delta}_{\alpha}\{X^{i},X^{j},\psi_{\delta\beta}\}+
\psi^{\dagger\alpha\beta}(\sigma^{ab})^{\delta}_{\alpha}\{X^{a},X^{b},\psi_{\delta\beta}\})
\nonumber\\
&&\
-fe^{\phi}\psi_{\dot{\alpha}\dot{\beta}}\psi^{\dagger\dot{\alpha}\dot{\beta}}
-\frac{2e^{-\phi}}{P^{+}}
(\psi_{\dot{\delta}\dot{\beta}}(\sigma^{ij})^{\dot{\delta}}_{\dot{\alpha}}
\{X^{i},X^{j},\psi^{\dagger\dot{\alpha}\dot{\beta}}\}
\nonumber\\
&&\
+\psi_{\dot{\delta}\dot{\beta}}(\sigma^{ab})^{\dot{\delta}}_{\dot{\alpha}}
\{X^{a},X^{b},\psi^{\dagger\dot{\alpha}\dot{\beta}}\})].
\end{eqnarray}
Here the $\psi$'s carry two spinor indices, corresponding to two
$SO(4)(SU(2)_{L}\times SU(2)_{R}$)'s Weyl spinor representations
respectively.

To do discrete light-cone quantization, the light-like direction
should be compactified.
 \begin{equation}
X^{-}=X^{-}+R_{-} .
\end{equation}
With this compactification, $P^{+}$ is quantized,
\begin{equation}
P^{+}=\frac{J}{R_{-}},
\end{equation}
where $J\in\mathbb{Z}$.

Next one has to quantize Nambu three brackets in order to have a
 matrix regularization of the
D$3$-brane Hamiltonian. It is well known that the quantization of
Nambu odd brackets is very difficult, because some important
properties of Nambu odd brackets will get lost if we apply the usual
way of quantization of Nambu even brackets. In \cite{jabatiny},
Sheikh-Jabbari proposed a way to quantize the Nambu odd bracket by
introducing a special constant matrix. For Nambu three brackets, the
constant matrix is $\mathcal {L}_{5}$, and the quantization is to
replace $\{F,G,K\}$ with $ J[F_{J\times J},G_{J\times J},K_{J\times
J},\mathcal {L}_{5}]$, where
\begin{eqnarray}
[F_{1},F_{2},F_{3},F_{4}]\equiv
\frac{1}{24}\epsilon^{i_{1}i_{2}i_{3}i_{4}}F_{i_{1}}F_{i_{2}}F_{i_{3}}F_{i_{4}}.
\end{eqnarray}
The $\mathcal {L}_{5}$ is made out of the direct product of unit
matrix $I_{4\times 4}$ and $SO(4)$ chirality operator $\Gamma^{5}$.
%It is associated with the representation of $Spin(4)$ and its
%definition can be found in \cite{jabatiny}.
In this paper, we only need one property of it. Namely there exist
matrices $X^{i}_{J\times J}, \ i=1,2,3,4$, which satisfy
\begin{eqnarray}
\label{fuzzy}
 [X^{i}_{J\times J},X^{j}_{J\times J},X^{k}_{J\times J},\mathcal
{L}_{5}]&=&-\frac{R^{2}}{J}\epsilon^{ijkl}X^{l}_{J\times J},
\nonumber\\
\sum_{i=1}^{4}(X^{i}_{J\times J})^{2}&=&R^{2}I_{J\times J}.
\end{eqnarray}
Recalling that  a three sphere of radius $R$ can be embedded in four
dimensional space as
\begin{eqnarray}
\{X^{i},X^{j},X^{k}\}&=&R^{2}\epsilon^{ijkl}X^{l},
\nonumber\\
\sum_{i=1}^{4}(X^{i})^{2}&=&R^{2},
\end{eqnarray}
we know that the matrices satisfying (\ref{fuzzy}) is actually a
realization of fuzzy three sphere $S^{3}_{F}$ of radius $R$. For the
explicit construction, please see \cite{jabatiny} and
\cite{Ramgoolam01}.

Now for the sector with $J$ units of light-cone momentum, we replace
$X^{I}, P^{I}, \psi_{\alpha\beta}, \psi_{\dot{\alpha}\dot{\beta}}$
with $J\times J$ matrices.
\begin{eqnarray}
X^{I}&\to& X^{I}_{J\times J}
\nonumber\\
P^{I}&\to& J\Pi^{I}_{J\times J}
\nonumber\\
\psi&\to&\sqrt{J}\psi_{J\times J}.
\end{eqnarray}
In the following part of this paper, we will denote
$X^{I}_{J\times J}, \Pi^{I}_{J\times J}, \psi_{J\times J}$ with
$X^{I}, \Pi^{I}, \psi$. We wish this will not bring any confusion.

With all these preparation, we can finish constructing the matrix
model for DLCQ of IIB string theory on background
(\ref{background}) by doing the replacements:
\begin{equation}
\frac{1}{P^{+}}\int d^{3}\sigma \to R_{-}\textrm{Tr}
\end{equation}
\begin{equation}
\{F,G,K\}\to J[F,G,K,\mathcal {L}_{5}].
\end{equation}
Then the matrix model Hamiltonian is
\begin{eqnarray} \label{conjecture}
H&=&R_{-} \textrm{Tr}
\{\frac{1}{2}\Pi^{2}_{I}+\frac{1}{2}\frac{\lambda}{R_{-}^{2}}X^{2}_{I}+\frac{e^{-2\phi}}{2\cdot3!}
[X^{I},X^{J},X^{K},\mathcal {L}_{5}][X^{I},X^{J},X^{K},\mathcal
{L}_{5}]
\nonumber\\
&&\qquad-\frac{f}{3!R_{-}}(\epsilon^{ijkl}X^{i}[X^{j},X^{k},X^{l},\mathcal
{L}_{5}]+\epsilon^{abcd}X^{a}[X^{b},X^{c},X^{d},\mathcal {L}_{5}])
\nonumber\\
&&\qquad+\frac{fe^{\phi}}{R_{-}}(\psi^{\dagger\alpha\beta}\psi_{\alpha\beta}
-\psi_{\dot{\alpha}\dot{\beta}}\psi^{\dagger\dot{\alpha}\dot{\beta}})
+2e^{-\phi}(\psi^{\dagger\alpha\beta}(\sigma^{ij})^{\delta}_{\alpha}[X^{i},X^{j},\psi_{\delta\beta},\mathcal
{L}_{5}]
\nonumber\\
&&\qquad+\psi^{\dagger\alpha\beta}(\sigma^{ab})^{\delta}_{\alpha}[X^{a},X^{b},\psi_{\delta\beta},\mathcal
{L}_{5}])
-2e^{-\phi}(\psi_{\dot{\delta}\dot{\beta}}(\sigma^{ij})^{\dot{\delta}}_
{\dot{\alpha}}[X^{i},X^{j},\psi^{\dagger\dot{\alpha}\dot{\beta}},\mathcal
{L}_{5}]
\nonumber\\
&&\qquad+\psi_{\dot{\delta}\dot{\beta}}(\sigma^{ab})^{\dot{\delta}}_
{\dot{\alpha}}[X^{a},X^{b},\psi^{\dagger\dot{\alpha}\dot{\beta}},\mathcal
{L}_{5}])\}
\end{eqnarray}
In this Hamiltonian, $\lambda,\phi,f$ depend on $\tau$, and satisfy
\begin{equation}
\lambda(\tau)=-\frac{1}{4}\phi''(\tau)+f^{2}(\tau)e^{2\phi(\tau)}.
\end{equation}

Now we have constructed a $(0+1)$-dimensional matrix model of type
IIB string theory on time-dependent background (\ref{background}).
%It is
%time-dependent just like the background. At first, let us
%analysis its symmetry.
The symmetry algebra of background (\ref{background}) is
$[so(4)\oplus so(4)]\oplus_{s}h(8)$, where $h(8)$ is
Heisenberg-type algebra.  %, and every generator of it contain
%$\partial_{x^{-}}$.
So DLCQ of string theory on this background should respect
$SO(4)\times SO(4)$ symmetry. From (\ref{conjecture}), it is easy to
see that our matrix model do have this symmetry.

Generically the matrix model has time-dependent mass-squared
$\lambda$ and the coupling constant is also time-dependent. In
this paper, we are interested in the null linear dilaton
background
\begin{equation}\label{philea} \phi=-c\tau,\ c>0.
\end{equation}
so that \begin{equation} \lambda=f^2 e^{2\phi}.\end{equation}

As the usual PP-wave case, the string theory limit could be
\begin{eqnarray}
J,R_{-}\to\infty, \hspace{5ex}\mbox{with $P^{+}$ being fixed.}
\end{eqnarray}
In this limit, our matrix model should describe light-cone type
IIB string theory on the background (\ref{background}).

Because our matrix model is time-dependent, it is difficult to find
a static solution. The unique static solution is the vacuum
solution:
\begin{equation}
X^{i}=X^{a}=\psi_{\alpha\beta}=\psi^{\dot{\alpha}\dot{\beta}}=0.
\end{equation}

Let us consider the fluctuations around the vacuum. Expanding the
Hamiltonian to the second order of the fluctuations around the
vacuum, we obtain
\begin{eqnarray}\label{secham}
H^{(2)}_{tri}&=&\textrm{Tr}[(\frac{R_{-}}{2}\Pi^{2}_{i}+\frac{\lambda}{2R_{-}}X^{2}_{i})
+(\frac{R_{-}}{2}\Pi^{2}_{a}+\frac{\lambda}{2R_{-}}X^{2}_{a})
\nonumber\\
&&\quad+fe^{\phi}\psi^{\dagger\alpha\beta}\psi_{\alpha\beta}
-fe^{\phi}\psi_{\dot{\alpha}\dot{\beta}}\psi^{\dagger\dot{\alpha}\dot{\beta}}],
\end{eqnarray}
where we use  $X^{i}, X^{a},
\psi_{\alpha\beta},\psi^{\dot{\alpha}\dot{\beta}}$ to denote the
fluctuations.

The lowest excited states around the vacuum are \begin{equation}
\textrm{Tr}(\sqrt{\frac{\lambda^{\frac{1}{2}}}{2R_{-}}}X^{I}-i\sqrt{\frac{R_{-}}{2\lambda^{\frac{1}{2}}}}\Pi_{I})\vert
0\rangle,\ \textrm{Tr}\psi^{\dagger\alpha\beta}\vert0\rangle,\
\textrm{Tr}\psi^{\dagger\dot{\alpha}\dot{\beta}}\vert0\rangle.
\end{equation}
The bosonic excitations have energy
\begin{equation}
E=\lambda^{\frac{1}{2}},
\end{equation}
which depends only on the background geometry. And the fermionic
excitations have energy
\begin{equation}
E=fe^{\phi},
\end{equation}
which depend on the background fluxes. If $\phi=ax^{+}+b,\ a,b$ are
constant, the first excited bosonic and fermionic states will have
the same energy. These energy are generically time-dependent except
a few special cases. One of such exception is the usual plane-wave
background, which has been discussed in \cite{jabatiny}. Another
special example is
\begin{eqnarray}\label{lamcon}
\phi&=&-c\tau,\ f=f_{0}e^{c\tau},
\end{eqnarray}
with $\lambda=f_{0}^{2}= const.$, so the first excited states have
the fixed energy.

The higher excited states can be discussed similarly. As the first
excited states,  their energy will be generically time-dependent.

Though we have computed the energy of the excited states, they are
only perturbational results in the matrix model. They are only
trustable if the coupling in the matrix model be very small. If we
define
\begin{eqnarray}
\widetilde{X}^{I}=\sqrt{\frac{1}{R_{-}}}X^{I},\hspace{3ex}\widetilde{\Pi}_{I}
=\sqrt{R_{-}}\Pi^{I},
\end{eqnarray}
the bosonic part of the matrix model Hamiltonian about this vacuum
is
\begin{eqnarray}\label{couple}
H_{bos}&=&
\textrm{Tr}\{\frac{1}{2}\widetilde{\Pi}^{2}_{I}+\frac{\lambda}{2}\widetilde{X}^{2}_{I}
\nonumber\\
&&\quad+\frac{1}{2\cdot3!}(R_{-}e^{-\frac{\phi}{2}})^{4}
[\widetilde{X}^{I},\widetilde{X}^{J},\widetilde{X}^{K},\mathcal
{L}_{5}][\widetilde{X}^{I},\widetilde{X}^{J},\widetilde{X}^{K},\mathcal
{L}_{5}]
\\
&&\quad-\frac{1}{3!}(R_{-}f^{\frac{1}{2}})^{2}(\epsilon^{ijkl}\widetilde{X}^{i}
[\widetilde{X}^{j},\widetilde{X}^{k},\widetilde{X}^{l},\mathcal
{L}_{5}]+\epsilon^{abcd}\widetilde{X}^{a}[\widetilde{X}^{b},\widetilde{X}^{c},\widetilde{X}^{d},\mathcal
{L}_{5}])\}\nonumber
\end{eqnarray}

The 't Hooft coupling in (\ref{couple}) is
\begin{equation}
g=JR_{-}^2e^{-\phi}=\frac{J^{3}e^{c\tau}}{(P^{+})^{2}}
\end{equation}
In the early time, $\tau\to -\infty$, the 't Hooft coupling will
be very weak. So we can trust the perturbation theory of matrix
model. %At the same time, the first excited states will be very
%heavy.
On the contrary, in the late time, when $\tau\to \infty$, the
perturbation matrix calculation will be not convincing, but the
perturbative string theory is well-defined then.

%In Jabbari's tiny graviton matrix model, the `trivial vacuum' is
%quite non-trivial. Jabbari conjecture that in his matrix model for
%DLCQ of string on plane-wave background, the
%$X^{i}=X^{a}=\psi_{\alpha\beta}=\psi^{\dot{\alpha}\dot{\beta}}=0$
%vacuum correspond to the vacuum of string $\sigma-$model on
%plane-wave background ignoring the string tension term. So spectrum
%of supergravity on plane-wave background \cite{jabarev} should be
%reproduced from second order Hamiltonian on trivial vacuum. In our
%case, we can compare our second order Hamiltonian (\ref{secham})
%with the $\sigma$-model Hamiltonian in \cite{chen}. If
%$\phi=-c\tau$, definition
For the case (\ref{couple}), defining
\begin{eqnarray}
a_{I}&=&
\sqrt{\frac{fe^{-c\tau}}{2R_{-}}}X^{I}+i\sqrt{\frac{R_{-}}{2fe^{-c\tau}}}\Pi_{I},
\nonumber\\
a^{\dagger}_{I}&=&
\sqrt{\frac{fe^{-c\tau}}{2R_{-}}}X^{I}-i\sqrt{\frac{R_{-}}{2fe^{-c\tau}}}\Pi_{I},
\end{eqnarray}
we obtain the normal-ordered free Hamiltonian
\begin{eqnarray}
 H^{(2)}_{tri}=\textrm{Tr}[fe^{-c\tau}(a^{\dagger}_{I}a_{I}+\psi^{\dagger\alpha\beta}\psi_{\alpha\beta}
+\psi^{\dagger\dot{\alpha}\dot{\beta}}\psi_{\dot{\alpha}\dot{\beta}})].
\end{eqnarray}
The zero-point energy is canceled between bosonic and fermionic
sectors in the background (\ref{background}). The free Hamiltonian
is the same as the one with only zero modes in perturbative string
vacuum. This suggests that in the string theory limit, the static
vacuum becomes the vacuum of string theory on plane-wave background
with linear null dilaton.

In the tiny graviton matrix model on plane-wave background, if one
takes the string theory limit, the 't Hooft coupling becomes very
large, which suggests that the closed string will appear as
non-perturbative objects of the model. In our case, the string
theory limit is more subtle. Since the background is time-dependent,
the matrix model and its 't Hooft coupling is also time-dependent.
Strictly speaking, the matrix model is only well-defined in the very
early time when the matrix coupling is small and the string coupling
is large,  and on the other hand the string theory is well-defined
in the very late time when the string coupling is small and the
matrix coupling is large. If we take the string theory limit at
fixed time where the string coupling is not very large, the 't Hooft
coupling becomes very large just like the plane-wave background
cases, and the vacuum in matrix model changes to closed
string vacuum.. %But
%if we take string theory limit with time evolution, the behavior
%of the 't Hooft coupling will depend on $f$.
%For the example (\ref{philea}),
%We should take late time limit with string theory limit so that we
%can obtain a weak coupling string theory.

\section{The fuzzy three sphere solutions}
In this section, we will discuss fuzzy three sphere solutions.  In
our time-dependent matrix model, there is no static fuzzy sphere
solution and the only possible fuzzy three sphere solutions are
time-dependent. Since our background keeps $SO(4)\times SO(4)$
symmetry, without losing generality, we take the following ansatz
\begin{eqnarray}
X^{a}&=&\psi_{\alpha\beta}=\psi_{\dot{\alpha}\dot{\beta}}=0,
\nonumber\\
X^{i}&=&S(\tau)X^{i}_{0},\nonumber\\
i&=&1,2,3,4,\ a=5,6,7,8.
\end{eqnarray}
Here $X^{i}_{0}$'s are constant $J\times J$ matrices satisfying
\begin{eqnarray}
[X^{i}_{0},X^{j}_{0},X^{k}_{0},\mathcal
{L}_{5}]&=&-\epsilon^{ijkl}X^{l}_{0},
\nonumber\\
\sum_{i=1}^{4}(X^{i}_{0})^{2}&=&J,
\end{eqnarray}
such that
\begin{eqnarray}
[X^{i},X^{j},X^{k},\mathcal
{L}_{5}]=-S^{2}(\tau)\epsilon^{ijkl}X^{l},
\nonumber\\
\sum_{i=1}^{4}(X^{i})^{2}=S^{2}(\tau)J.
\end{eqnarray}

With this ansatz, the equation of motion for $X^{i}$ is
\begin{eqnarray}
\label{equation} \ddot{X}^{i}+\lambda
X^{i}-\frac{4R_{-}f}{3!}\epsilon^{ijkl}[X^{j},X^{k},X^{l},\mathcal
{L}_{5}]
\nonumber\\
-\frac{R_{-}^{2}e^{-2\phi}}{2}[[X^{i},X^{j},X^{k},\mathcal
{L}_{5}],X^{j},X^{k},\mathcal {L}_{5}]=0,
\end{eqnarray}
which gives the equation for $S(\tau)$,
\begin{eqnarray}\label{fuzzyeq}
\ddot{S}(\tau)+\lambda
S-4R_{-}fS^{3}(\tau)+3R_{-}^{2}e^{-2\phi}S^{5}(\tau)=0.
\end{eqnarray}
The linear term of the above equation stems from the mass-squared
term in the matrix model, which is determined by the background
geometry. The other two terms, cubic one and quintic one, are from
the interaction terms in the matrix model. There are two
parameters $\phi(\tau)$ and $f(\tau)$ in the equation. In this
article, we discuss two cases whose corresponding perturbative
string have been studied in \cite{chen}:
 \begin{itemize}
 \item Case 1:
$ \phi=-c\tau,\ f(\tau)=f_{0}e^{c\tau},\ c>0,\ f_{0}=const$;
 \item Case 2: $\phi=-c\tau,\  f(\tau)=f_{0},\ c>0,\
f_{0}=const.$
\end{itemize}

In both cases, $\phi=-c\tau$, so (\ref{fuzzyeq}) is
\begin{equation}\label{seq}
\ddot{S}(\tau)+f^{2}e^{-2c\tau}S-4R_{-}fS^{3}(\tau)+3R_{-}^{2}e^{2c\tau}S^{5}(\tau)=0.
\end{equation}
It is very difficult to obtain analytic solutions of this
equation. Using the numerical method to discuss the solution and
combining the analysis of  the early time and late time asymptotic
behavior of the solution, we will obtain a good qualitative
understanding of
 the solution.

For the first case, $\lambda=f_{0}^{2}$ is a constant. The
equation for $S(\tau)$ is
\begin{equation}
\label{equationfe}
\ddot{S}(\tau)+f^{2}_{0}S(\tau)-4R_{-}f_{0}e^{c\tau}S^{3}(\tau)+3R_{-}^{2}e^{2c\tau}S^{5}(\tau)=0.
\end{equation}
The solutions  depend on three parameters, $R_{-},  f_{0}, c$. For
a definite discussion, at first, we let
\begin{eqnarray} \label{set1}
 R_{-}=2,\hspace{3ex} f_{0}=1, \hspace{3ex} c=1.
\end{eqnarray}
%Now we can  try to look for numerical solutions.
%\begin{widetext}
\begin{figure}[htb]
\begin{center}
%\scalebox{2.0}{
\includegraphics[width=0.5\textwidth]{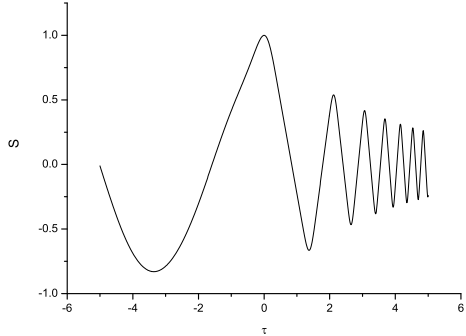}
%}height=0.25\textheight,
\end{center}
\caption{The solution of Case 1 when $\tau\in [-5,5]$ and $R_{-}=2,
f_{0}=1, c=1$. At late time, this solution vibrate faster with a
smaller amplitude. In fact, its amplitude will shrink to zero very
quickly. } \label{fig1}
\end{figure}
%\end{widetext}

The figure \ref{fig1} is for the numerical solution between
$\tau=-5$ and $\tau=5$ if we set initial conditions
\begin{equation}
S(0)=1,\hspace{3ex} \dot{S}(0)=0.
\end{equation}
From this figure, we can read out the behavior of $S(\tau)$ in
$[-5,5]$. When $\tau$  becomes larger, in other words, when $g_{s}$
becomes weaker, $S(\tau)$ vibrates more rapidly, and the amplitude
becomes smaller. The figure can not tell us the whole story, because
when $\tau\gg 1$, $S(\tau)$ vibrates so rapidly that the computer
can not give right numerical solutions. Fortunately, when $|\tau|$
is large, we can analyze the solutions from their asymptotic
behaviors.

At early time, if we take $\tau\to -\infty$ limit, the equation is
dominated by the linear term and is approximated to be
\begin{equation}
\label{app}
 \ddot{S}(\tau)+f^{2}_{0}S(\tau)=0,
\end{equation}
which has the solution
\begin{eqnarray}
S(\tau)=A\textrm{sin}(f_{0}\tau+B).
\end{eqnarray}
$A,B$ are constants which  depend on initial conditions. So at
early time, fuzzy three sphere is in a stable vibrating state.

At late time, when $\tau\to\infty$, the equation
(\ref{equationfe}) is dominated by the interaction term and can be
approximated to be
\begin{equation}
\label{simplefe}
 \ddot{S}(\tau)+3R_{-}^{2}e^{2c\tau}S^{5}(\tau)=0.
\end{equation}
So at very late time, $S(\tau)$ will vibrate very rapidly around
zero. The amplitude will also decay to zero very fast, at the rate
of $e^{-\frac{c}{2}\tau}$.  This behavior is expected. From matrix
model point of view, at late time, the coupling is stronger and
the nonperturbative effect becomes essential and the fundamental
closed string degree of freedom becomes important so the fuzzy
effect is quite weak. %When $g_{s}$ became weaker, the fuzzy effect
%should also be weaker.

Therefore, we have the following  picture. A fuzzy three sphere,
 vibrates in a stable rhythm in the early time. But after a
long time evolution, at late time, the non-Abelian degrees of
freedom of the matrix model become less important and the
fundamental string degrees of freedom take it over.

\begin{center}
\begin{figure}[htb]
%\scalebox{2.0}{
\includegraphics[width=0.5\textwidth]{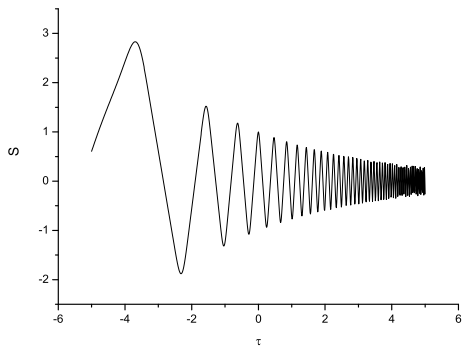}
%}height=0.25\textheight,
\caption{The solution of Case 1 when $\tau\in [-5,5]$ and $R_{-}=10,
f_{0}=1, c=1$. At late time, it vibrates more rapidly than the one
in $R_{-}=2$ case which is plotted in figure
\ref{fig1}.}\label{fig2}
\end{figure}
\end{center}

Changing parameters in the equation will not change the whole
picture but can postpone or accelerate the solutions to reach their
asymptotic behaviors. If we change the setting, let
\begin{equation}
\label{set2}
 R_{-}=10,\hspace{3ex} f_{0}=1,\hspace{3ex} c=1,
\end{equation}
and still set the initial conditions
\begin{equation}
S(0)=1,\hspace{3ex} \dot{S}(0)=0,
\end{equation}
we obtain the figure \ref{fig2}.

%\begin{widetext}
%\end{widetext}
Comparing the figures (\ref{fig1}) and (\ref{fig2}), we find that at
late time, $S(\tau)$ vibrates more rapidly when $R_{-}$ becomes
larger. This is because the coefficient of the quintic term
 is larger, which suggests a stronger interaction and leads to faster vibration.
  In other words, a larger $R_{-}$ makes fuzzy effect
weaker. If we let $f_{0}$ be larger, from (\ref{equationfe}), we can
deduce that the fuzzy three sphere will shrink more slowly.

\begin{figure}[htb]
\begin{center}
%\scalebox{2.0}{
\includegraphics[width=0.5\textwidth]{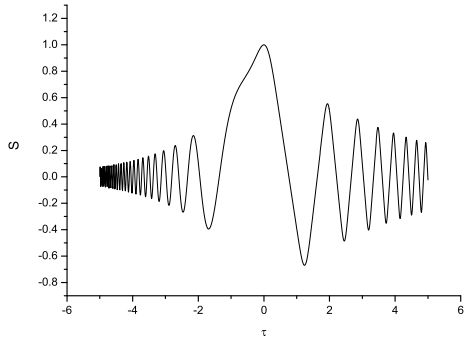}
%}height=0.25\textheight,
 \caption{The solution of Case 2 when $\tau\in [-5,5]$ and
$R_{-}=2, f_{0}=1, c=1$. At late time, its asymptotic behavior is
similar to the asymptotic behavior of solutions in Case 1. At early
time, the solution has a similar asymptotic behavior to its late
time asymptotic behavior. }
 \label{fig3}
\end{center}
\end{figure}

For the second case, $\lambda=f_{0}^{2}e^{-2c\tau}$, the equation
(\ref{fuzzyeq}) is
\begin{equation}
\label{equationf}
\ddot{S}(\tau)+f^{2}_{0}e^{-2c\tau}S-4R_{-}f_{0}S^{3}(\tau)+3R_{-}^{2}e^{2c\tau}S^{5}(\tau)=0.
\end{equation}
At first, we set $R_{-}=2, f_{0}=1, c=1$ just like (\ref{set1}). If
we input
\begin{equation}\label{incon}
S(0)=1,\dot{S}(0)=0,
\end{equation}
we have the figure \ref{fig3} for $\tau\in[-5,5]$.

%\begin{widetext}
%\end{widetext}
From this figure, we find that the late time asymptotic behavior of
this solution is very similar to the one in Case 1. But the early
time asymptotic behavior is very different. Just like Case 1, we can
not obtain conclusions from figure when $|\tau|\gg 1$. So we must
analyze the asymptotic behaviors of  the solutions.

At late time, when $\tau\to\infty$, the equation (\ref{equationf})
is approximated to be
\begin{eqnarray}
\label{simplef}
 \ddot{S}(\tau)+3R_{-}^{2}e^{2c\tau}S^{5}(\tau)=0.
\end{eqnarray}
It is the same equation as (\ref{simplefe}). So the asymptotic
behaviors of solutions in these two cases at late time are the
same. We know at late time, $\lambda,g_{s}\to 0$, $f=f_{0}$ still
is constant. So the decay of fuzzy three sphere is a natural
phenomenon because the background now is nearly flat and the
string coupling is weak and the matrix coupling is strong.

At very early time, the equation (\ref{equationf}) is approximated
to be
\begin{eqnarray}
\label{simplef2} \ddot{S}(\tau)+f^{2}_{0}e^{-2\tau}S(\tau)=0.
\end{eqnarray}
This equation tell us that  when $\tau\to -\infty$, the asymptotic
behavior of the solution of (\ref{equationf}) will be very similar
to its late time one. So the behavior shown in figure (\ref{fig3})
is inevitable. To understand these behavior, let us come back to
equation (\ref{equation}). The most important term dominating the
early time behavior, is $\lambda X^{i}$, which depends on the
background geometry. Since $\lim_{\tau\to -\infty}\lambda=\infty $
in this case,  the large spatial extension reduce fuzzy effect. This
phenomenon happens also in the matrix models on other time-dependent
backgrounds\cite{chen2,das05}.
%A possible reason is that at early time, gravity is
%too strong, so that we can not believe the solutions of classic
%equation of motion.

Nevertheless, we can read out the whole behavior of solutions from
the figure \ref{fig3}, (\ref{simplef}) and (\ref{simplef2}). If
there is a very small fuzzy three sphere at early time, it will grow
up as time goes by. And after it reaches its maximum, it becomes
smaller and smaller. At late time, it vibrates fiercely and decays
out.

\begin{center}
\begin{figure}[htb]
%\scalebox{2.0}{
\includegraphics[width=0.5\textwidth]{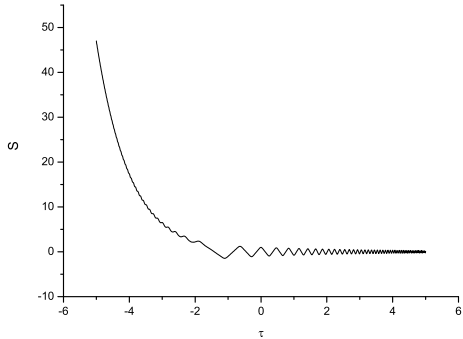}
%}height=0.25\textheight,
\caption{The solution of Case 2 when $\tau\in [-5,5]$ and $R_{-}=10,
f_{0}=1, c=1$. At late time, this solution shrinks to zero much
faster than the one when we set $R_{-}=2, f_{0}=1, c=1$. At early
time, though its full asymptotic behavior can not be seen from the
figure because we do not plot the figure of the solution when
$\tau<-5$, we can confirm that its asymptotic behavior is similar to
$R_{-}=2$ case through analysis in  this section.} \label{fig4}
\end{figure}
\end{center}

We can also analyze the effect of changing parameters. If we use the
setting
\begin{eqnarray}
 R_{-}=10,\hspace{3ex} f_{0}=1,\hspace{3ex} c=1,
\end{eqnarray}
we get the figure \ref{fig4} for the solution between $\tau=-5$ and
$\tau=5$ with the initial conditions (\ref{incon}).

%\begin{widetext}
%\end{widetext}
As before, a larger $R_{-}$ makes the solution vibrate more rapidly
when $\tau\to\infty$, and makes the solution vibrate more slowly
when $\tau\to -\infty$ (Since we only plot the figure when $\tau\in
[-5,5]$, the early time asymptotic behavior of the solution is not
evident in the figure 4). If we let $f_{0}$ take a larger value,
however, we will find contrary effect.

Before ending our study of fuzzy three sphere, it would be
interesting to discuss how the tiny graviton scale changes with
time. In the appendix, we investigate  the giant gravitons in the
background (\ref{background}). We learn that the size of giant
graviton is changing with time:
\begin{equation}
\frac{d^2r}{d\tau^2}+\lambda r-4f(\frac{\mu }{P_-})r^3+3(\frac{\mu
}{P_-})^2r^5e^{-2\phi(\tau)}=0, \label{ggradius}
\end{equation}
where $\mu=2\pi^2 T_3$ with $T_3$ being the tension of D3-brane
and $P_-$ is conserved lightcone momentum. In the case
$\phi=\mbox{constant}, \lambda=f^2_0$, there exist static giant
gravitons with radius $r_0^2=f_0\frac{P_-}{\mu}$ and
$r_0^2=3f_0\frac{P_-}{\mu}$. The former one is the well-studied
stable giant graviton keeping half supersymmetries. The latter one
is not stable. From this, one can read out the size of tiny
graviton which carry one unit of lightcone momentum,
\begin{equation}
r_{\mbox{\tiny tiny}}=\frac{l^2_{\mbox{\tiny pl}}}{r_{\mbox{\tiny
AdS}}}.
\end{equation}
For the case $\phi$ being not a constant, after identifying
$P_-=\mu/{R_-}$, we can find that the equation (\ref{ggradius})
takes the same form as the equation (\ref{fuzzyeq}). This shows
that the scale of tiny graviton changes with $\tau$ in the same
way as the size of fuzzy three sphere. This is not a surprise
since the tiny graviton is the fundamental parton of our matrix
model. This fact could be understood as following: the fuzzy three
sphere could be taken as the blow-up of the tiny gravitons,
therefore its size is changing with the evolution of the scale of
tiny graviton.

Actually the fact that the dynamics of fuzzy three sphere is in
perfect agreement with the dynamics of giant graviton in the
background (\ref{background}) gives a consistent check that our
matrix model on time-dependent background is correct.

\section{Conclusions and discussions}

In this paper, we constructed a matrix model for type IIB
superstring theory on a plane-wave background with null dilaton
(\ref{background}). Our matrix model is time-dependent, have much
different character from the tiny matrix model on plane-wave
background. One of the differences is that the vacuum solution is
the unique static solution. The spectrum about this vacuum are
time-dependent in most generic cases. We compared our second order
Hamiltonian of trivial vacuum to Hamiltonian of string
$\sigma$-model on the linear null dilaton background, and found that
the spectrum of perturbation of background in string theory could be
produced from our second order Hamiltonian.

We also investigated the fuzzy three-sphere solutions in our matrix
model and found that they must be time-dependent. The solutions
depend on the background geometry and flux.  In this paper, we
discuss two cases with linear null dilaton. One is
$\phi={-c\tau},f(\tau)=f_{0}e^{c\tau},f_{0}=constant$. The other one
is $\phi={-c\tau},f(\tau)=f_{0}=constant$. For these two cases, we
could not obtain analytic solutions yet, but we investigated the
solutions numerically  and also analyzed their early and late time
asymptotic behaviors. The results are interesting. In the first
case, a fuzzy sphere vibrates stably at early time, but at late time
it vibrates much more rapidly, then decays out. In the second case,
at early time a very small fast-vibrating fuzzy sphere appears and
grows up gradually to reach its steady maximum and then decays
through rapid vibration at late time. The picture can be understood
as following. The early time behavior of the fuzzy three-sphere is
determined by the background geometry completely, so in the second
case, the fuzzy degrees of freedom is diluted. On the other hand,
the late time asymptotic behavior is determined by the interaction
terms in the matrix model. The very strong interaction at the late
time requires the nonperturbative degrees of freedom to replace the
matrix degrees of freedom. This explain why the fuzzy sphere always
decays out at the late time.

The matrix model constructed here is well-defined at the very early
time where the coupling of the interaction is weak. It is dual to
the late time perturbative string description. In other words, the
closed string perturbative string will appear as the nonperturbative
objects in the matrix model, and vice versa. The dual pictures give
a complete description of the string theory in the linear null
dilaton background.

In the matrix model on plane-wave background\cite{jabatiny}, the
spectrum around fuzzy sphere vacuum contain photon states, which are
the gauge field on the giant graviton in the string theory limit. In
our case, fuzzy sphere solutions are too complex and time-dependent.
We did not go to similar discussions in this paper. We hope we can
discuss these issues in the future.

One essential ingredient in the construction of tiny graviton matrix
model is the existence of  Nambu three-bracket in the D3-brane
action, which allow us to do matrix regularization to obtain a
matrix model. On the other hand, the Nambu three-bracket is a
realization of infinite dimensional 3-algebra, which plays a central
role in the recent study of multiple membrane theory by
Bagger-Lambert\cite{bl1,bl2} and Gustavsson\cite{gus}.  In
\cite{ho1,pkt}, it was pointed out that  the space of functions on
three dimensional manifold, with Nambu three bracket could be
identified as an infinite dimensional 3-algebra. The Nambu three
bracket is the realization of three-bracket of this infinite
dimensional 3-algebra. Especially, it was proposed in \cite{Gomis08}
that the deformed BLG-theory on $R\times T^2$ gives tiny graviton
matrix theory of Type IIB plane-wave.\footnote{The relation between
mass deformed M2 theory on $T^2$ and IIB DLCQ string on plane-wave
 was first pointed out and identified in \cite{LLM}.}
Very recently, in \cite{jabarel}, the authors used the similar way
to \cite{jabatiny} as an matrix realization of ``relaxed
three-algebra ", which is a `3-algebra' with relaxed closure and
fundamental identity conditions. The relaxed three-algebra is a
linear space of matrices with a relaxed three bracket, which is
defined as $[T^{a},T^{b},T^{c},T^{-}]$ with $T^{-}$ being a special
fixed constant matrix. Using relaxed three-algebra, they tried to
construct a Hermitian model to describe $n$ (which is a general
number) membranes systems. These discoveries suggest a deep but not
clear relation between multi-membrane theory and tiny graviton
matrix model. It would be important to understand this issue better.
And it is also interesting to discuss the multiple membrane in a
time-dependent background\cite{Blau2}.

\section*{Acknowledgments}

The work was partially supported by NSFC Grant No. 10535060,
10775002, and NKBRPC (No. 2006CB805905). C would like to thank TIFR
for the hospitality, where the project was finishing.

\section*{Appendix: Giant gravitons}

In this appendix, we would like to study the giant gravitons in
the background (\ref{background}). Let us introduce the polar
coordinates $(r,\theta, \phi, \varphi)$ in the plane defined by
the Cartesian coordinates $x_i, i=1,2,3,4$. In polar coordinates,
the 5-form flux
\begin{equation}
(F_5)_{+1234}=(F_5)_{+r\theta\phi\varphi}=4f(x^+)r^3\sin^2\theta\sin\phi
\end{equation}
with the relevant nonvanishing components of 4-form potential
\begin{equation}
C_{+\theta\phi\varphi}=f(x^+)r^4\sin^2\theta\sin\phi.
\end{equation}
Here we just focus on the case that the giant graviton being blown
up in $x_i, i=1,2,3,4$. Obviously the case that the giant graviton
being blown up in $x_i, i=5,6,7,8$ could be studied similarly.

The giant gravitons correspond to spherical D3-brane embedded in
the background consistently. The D3-brane is wrapped around $S^3$
characterized by $\theta, \phi, \varphi$. This allows us to
identify the worldvolume coordinates of D3-branes to be $\tau,
\theta, \phi$ and $\varphi$. To simplify the discussion, we can
choose the gauge $X^+=\tau$ such that the only nontrivial
embedding is $r$ depending on $\tau$. After this truncation, the
action of the D3-brane reduce to that of a point particle:
 \begin{eqnarray}
 S&=&\mu \int d\tau [-e^{-\phi(\tau)}r^3(\tau)\sqrt{-G_{AB}\partial_\tau
 X^A\partial_\tau X^B}+f(\tau)r^4(\tau)],
 \end{eqnarray}
 where $\mu=2\pi^2 T_3$.\footnote{In our construction of matrix
 model, we start from a D3-brane action with $T_3=1$.}
As we have seen, the translation along $x^-$ is still an isometry
so its corresponding canonical momentum is a conserved quantity.
And the canonical Hamiltonian is just $H=-P_+$, which is of the
form
\begin{equation}
H=\frac{P_r^2}{2P_-}+\frac{(\mu r^3
e^{-\phi})^2}{2P_-}+\frac{\lambda r^2 P_-}{2}-\mu f r^4,
\end{equation}
where $P_r$ is the canonical momentum of $r$. The Hamiltonian
describes a particle with mass $P_-$ moving in a time-dependent
potential. The $r$ is actually the size of the giant graviton,
which is determined by the equation of motion
\begin{equation}
\frac{d^2r}{d\tau^2}+\lambda r-4f\frac{\mu}{P_-}r^3+3(\frac{\mu
}{P_-})^2e^{-2\phi}r^5=0.
\end{equation}
This equation is the same as (\ref{fuzzyeq}) after identification
$R_-=\mu/P_-$. Therefore the dynamics of giant graviton is in
exact agreement with fuzzy three sphere.

\end{document}